\documentclass[12pt]{article}
\usepackage{graphicx}
\usepackage{color}
\usepackage{epsfig}

\def\eqint{{\ \lower-1pt\vbox{\hbox{\rlap{$?$}\lower4pt\vbox{\hbox{$=$}}}}\ }} 

\def\siml{{\ \lower-1.2pt\vbox{\hbox{\rlap{$<$}\lower6pt\vbox{\hbox{$\sim$}}}}\ }}

\newcommand{\als}{\alpha_s}
\newcommand{\MS}{\overline{\rm MS}}
\newcommand{\RS}{\rm RS}

\def\lQ{\Lambda_{\rm QCD}}

\newcommand{\be}{\begin{equation}}
\newcommand{\ee}{\end{equation}}
\newcommand{\bea}{\begin{eqnarray}}
\newcommand{\eea}{\end{eqnarray}}

\newcommand{\nn}{\nonumber}
\newcommand{\al}{\alpha}

\def\pbnr{}
\def\speaker{Antonio Pineda}
\def\onbehalfof{}
\def\title{Radiative transitions in pNRQCD}
\def\affiliation{Grup de F\'\i sica Te\`orica\\
Universitat Aut\`onoma de Barcelona, Spain}
\def\support{This work was partially supported by the Spanish 
grants FPA2010-16963 and FPA2011-25948, and by the Catalan grant SGR2009-00894.
}

\newcommand\pubnumber{\pbnr}
\newcommand\pubdate{\today}
\def\Title#1{\begin{center} {\Large #1 } \end{center}}
\def\Author#1{\begin{center}{ \sc #1} \end{center}}

\newcommand{\OnBehalf}[1]{\sbox0{#1}\ifdim\wd0=0pt
        {}
	\else
	{\\on behalf of #1}
	\fi}
\newcommand{\SupportedBy}[1]{\sbox0{#1}\ifdim\wd0=0pt
        {}
	\else
	{\footnote{#1}}
	\fi}
\def\Address#1{\begin{center}{ \it #1} \end{center}}

\newcommand\pubblock{\includegraphics[width=5cm]{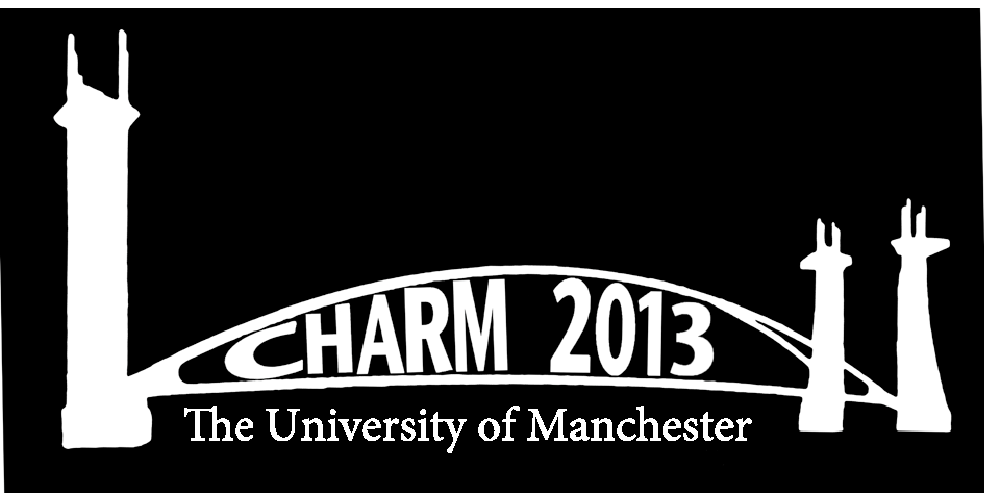}\hfill{\begin{tabular}{l} \pubnumber\\
         \pubdate  \end{tabular}}}
\newenvironment{Abstract}{\begin{quotation}  }{\end{quotation}}
\newenvironment{Presented}{\begin{quotation} \begin{center} 
             PRESENTED AT\end{center}\bigskip 
      \begin{center}\begin{large}}{\end{large}\end{center} \end{quotation}}

\def\venue{The 6$^{th}$ International Workshop on Charm Physics\\
(CHARM 2013)\\
Manchester, UK,  31 August -- 4 September, 2013}

\def\beq{\begin{equation}}
\def\eeq#1{\label{#1}\end{equation}}
\def\eeqn{\end{equation}}

\def\beqa{\begin{eqnarray}}
\def\eeqa#1{\label{#1}\end{eqnarray}}
\def\eeqan{\end{eqnarray}}

\let\bar=\overbar

\def\Dslash{\not{\hbox{\kern-4pt $D$}}}
\def\dslash{\not{\hbox{\kern-2pt $\del$}}}

\def\ee{e^+e^-}

\def\msb{{\bar{\ssstyle M \kern -1pt S}}}

\begin{document}
\begin{titlepage}
\pubblock

\vfill
\Title{\title}
\vfill
\Author{\speaker\SupportedBy{\support}\OnBehalf{\onbehalfof}}
\Address{\affiliation}
\vfill
\begin{Abstract}
We review recent model independent determinations of the radiative transitions of heavy quarkonium obtained using potential NRQCD.
\end{Abstract}
\vfill
\begin{Presented}
\venue
\end{Presented}
\vfill
\end{titlepage}
\def\thefootnote{\fnsymbol{footnote}}
\setcounter{footnote}{0}
%

\section{Introduction}

Heavy quarkonium states below threshold can be studied in a model independent way using 
pNRQCD~\cite{Pineda:1997bj}
(for some reviews see Refs.~\cite{Brambilla:2004jw,Pineda:2011dg}). pNRQCD is an effective field theory 
directly derived from
QCD. It efficiently disentangles the dynamics of the 
heavy quarks from the dynamics of the light degrees of freedom.  
It profits from the fact that the dynamics of the bound state system is   
characterized by, at least, three widely separated scales: hard (the
mass $m$ of the heavy quarks), soft (the relative momentum $|{\vec p}|
\sim mv \ll m$ of the heavy-quark--antiquark pair in the center-of-mass frame), and ultrasoft (the typical kinetic energy $E \sim
mv^2$ of the heavy quark in the bound state system).

The specific construction details of pNRQCD are slightly different depending on the relative size between the soft and the 
$\lQ$ scale. Two main situations are distinguished, namely, the weak-coupling \cite{Pineda:1997bj,Brambilla:1999xf} ($mv \gg \lQ$)  and the strong-coupling \cite{Brambilla:2000gk} ($mv \simeq \lQ$) versions of pNRQCD.
One major difference between them is that in the former the potential can be computed in perturbation theory unlike in the latter. 

In Ref. \cite{Brambilla:2005zw} the allowed ($n=n'$) and hindered ($n\not= n'$) magnetic dipole (M1) transitions between low-lying heavy quarkonium states 
were studied with pNRQCD in the strict weak-coupling limit. The authors of that work also performed a 
detailed comparison of the effective field theory and potential model (see Refs.~\cite{Voloshin:2007dx,Eichten:2007qx}
for some reviews) results.
The transitions considered were the following: 
$J/\psi(1S) \to \eta_c(1S)\,\gamma$, 
$\Upsilon(1S) \to \eta_b(1S)\,\gamma$, $\Upsilon(2S) \to \eta_b(2S)\,\gamma$, $\Upsilon(2S) \to \eta_b(1S)\,\gamma$,
$\eta_b(2S)\to\Upsilon(1S)\,\gamma$, $h_b(1P) \to \chi_{b0,1}(1P)\,\gamma$ and $\chi_{b2}(1P) \to h_b(1P) \,\gamma$. Large errors were assigned to the pure ground state observables, especially for charmonium, whereas disagreement with experimental bounds (at that time) was found for the hindered transition $\Upsilon(2S) \to \eta_b(1S)\,\gamma$. 

The perturbative expansion in the weak coupling version of pNRQCD was rearranged in Ref. \cite{Kiyo:2010jm} to improve its convergence. 
In this new expansion scheme the static potential 
(approximated by a polynomial of 
order $N+1$  in 
powers of $\als$) was exactly included in the leading order Hamiltonian ($C_f=(N_c^2-1)/(2N_c)$, $C_A=N_c$): 
\begin{eqnarray}
\label{H0}
H^{(0)}&\equiv& -\frac{{\bf \nabla}^2}{m}+V^{(N)}(r), \qquad {\rm and} \qquad H^{(0)}\phi_{nl}({\vec r})=E_{nl}\phi_{nl}({\vec r})
\,;
\\
V^{(N)}(r)
&=&
 -\frac{C_f\,\alpha_s(\nu)}{r}\,
\bigg\{1+\sum_{n=1}^{N}
\left(\frac{\alpha_s(\nu)}{4\pi}\right)^n a_n(\nu;r)\bigg\}
\,.
\label{VSDfo}
\end{eqnarray}
Thus, the natural expansion parameter, $\al_{V^{(N)}} \sim v$ is used. We expect (and find) convergence in $N$. In order to be so, one has to carefully implement the renormalon cancellation order by order in $\als$. This is particularly important if one tries to resum the soft logarithms in $V^{(N)}(r)$ setting $\nu=1/r$. This typically accelerates the convergence and makes the residual scale dependence smaller. 
We refer to \cite{Kiyo:2010jm,Pineda:2013lta} for details. The relativistic corrections are suppressed by powers of $v \equiv \sqrt{\langle p^2\rangle/m^2}$.
They depend on $N$, but the dependence is much smaller than their typical size, allowing a meaningful determination of them.

In Ref.\cite{Pineda:2013lta} this improved expansion scheme was applied to the M1 radiative transitions. 
The precision reached was $k_{\gamma}^3/m^2\times{\cal O}(\als^2,v^2)$ and $k_{\gamma}^3/m^2\times{\cal O}(v^4)$ for the allowed and forbidden transitions respectively, where $k_{\gamma}$ is the photon energy. Large hard logarithms (associated with the heavy quark mass) were also resummed. 
The effect of the new power counting was large and the exact treatment of the soft logarithms of the static potential made the factorization scale dependence much smaller. 
The convergence for the $b\bar b$ ground state was quite good. This allowed giving a solid prediction for the $\Upsilon(1S) \to \eta_b(1S)\,\gamma$ transition with small errors. 
The convergence was also quite reasonable for the $c\bar c$ ground state and the $b\bar b$ $1P$ state. For all of them solid predictions were also obtained. For the $J/\psi(1S)\to \eta_c(1S)\,\gamma$ transition the central value was significantly different from the one obtained in Ref. \cite{Brambilla:2005zw}, though perfectly compatible within errors. For the $2S$ decays the situation was less conclusive. Whereas for the $\Upsilon(2S) \to \eta_b(2S)\,\gamma$ decay there was not convergence, the previous disagreement with experiment for the hindered transition $\Upsilon(2S) \to \eta_b(1S)\,\gamma$ faded away with the new expansion scheme. 

Theoretical expressions for the E1 transitions of heavy quarkonium in pNRQCD have been obtained in 
\cite{Brambilla:2012be} both in the weak and strong coupling limit.  
A phenomenological analysis of these results was done in \cite{Pietrulewicz:2013ct}. Unlike 
for the M1 transitions, in the weak coupling limit the first non-perturbative effects enter at ${\cal O}(v^2)$ 
due to octet effects. For the strong coupling limit the result depends on wave functions and expectation values of operators obtained with non-perturbative potentials. In both cases, the treatment of the non-perturbative effects 
introduces larger uncertanties than for the M1 transitions.

\begin{figure}[!t]
\begin{center}
\hspace{0.50cm}
\epsfig{figure=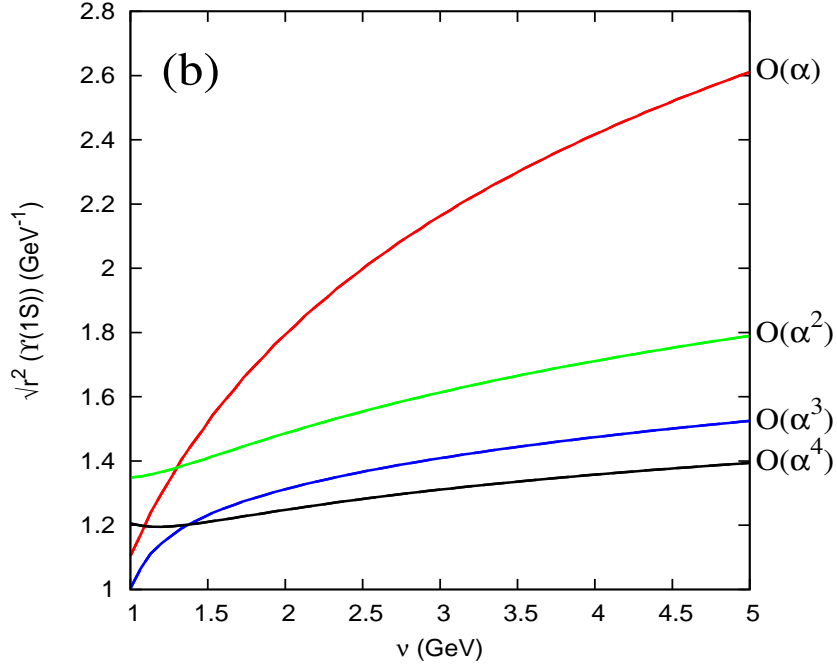,height=8.85cm,width=11.00cm}
\caption{\label{fig:botn1} \it Plot of $\sqrt{\langle r^2 \rangle_{10}}$ of the 
bottomonium ground state using the static potential $V_{\RS'}^{(N)}$ at different orders in perturbation theory: $N=0,1,2,3$.}
\end{center}
\end{figure}

\section{M1 transitions}

The allowed M1 radiative transitions depend on $\langle { p}^2 \rangle_{nl}$ and, at higher orders, on other expectation values such as 
$\langle r^2 \rangle_{nl}$. Studying them gives a hint of the applicability of the weak-coupling version of pNRQCD to those states, and a very nice check of the renormalon dominance picture. The matrix elements 
$\langle { p}^2 \rangle_{nl}$ and $\langle r^2 \rangle_{nl}$ of the charmonium ground state and of the $n=1,2$ bottomonium states were computed in \cite{Pineda:2013lta}. 

The electromagnetic radius was nicely convergent in all cases. Therefore, one can talk of the typical (electromagnetic) radius of these bound states. We illustrate this behavior for the radius of bottomonium ground state in Fig. \ref{fig:botn1}. The kinetic energy is also (though typically less than the radius) convergent,
except for the $2S$ state. Then, one can also define a $v^2_{nl}\equiv \langle p^2 \rangle_{nl}/m^2 $ for those states. 
These numbers are shown in Table \ref{tab:v2}.  
These numbers can be taken as estimates of the typical radius of the bound state system and of the typical velocity of the 
heavy quarks inside the bound state. For the $b\bar b$ ground state the result is very stable under scale variations; for the 
charm ground state and for the bottomonium $P$-wave the scale dependence is bigger. 

\begin{table}[htb]
\begin{center}
$$
\begin{array}{|l|c|c|c|c|}
\hline
   &b\bar b(1S)&c\bar c(1S)&b\bar b(1P)&b\bar b(2S)\\
\hline
v & 0.26 & 0.43 & 0.25 & 0.24 \\
\hline
\sqrt{\langle r^2 \rangle} ({\rm GeV^{-1}})& 1.2 & 2.2 & 2.1 & 2.9\\
\hline
\end{array}
$$ 
\end{center}
\caption{\label{tab:v2}\it Estimates for $v \equiv \sqrt{\langle {p}^2 \rangle/m^2}$ and $\sqrt{\langle r^2 \rangle}$ for the heavy quarkonium states. For the $b\bar b(2S)$ state the number given for $v$ was quite uncertain.}
\end{table}

In Ref. \cite{Pineda:2013lta} the following expressions were used for the allowed transitions
\bea
\label{nS=nS}
\Gamma(n^{3}S_{1}\to n^{1}S_{0}\gamma) &=& \frac{4}{3}\alpha e_Q^{2}
\frac{k_{\gamma}^{3}}{m^{2}}
\left[{\color{black}(1+\kappa_{})^2} - \frac{5}{3} 
\frac{{\color{black}\langle {p}^{\,2} \rangle_{n}}}{m^{2}}
\right], \\
\Gamma(n^{3}P_{J}\to n^{1}P_{1}\gamma) &
=&
\frac{3\Gamma(n^{1}P_{1}\to n^{3}P_{J}\gamma)}{2J+1}
= 
\frac{4}{3}\alpha e_Q^{2}
\frac{k_{\gamma}^{3}}{m^{2}}
\left[{\color{black}(1+\kappa_{})^2} - d_{J} 
\frac{{\color{black}\langle {p}^{2} \rangle_{n1}}}{m^{2}}
\right], 
\label{nP=nP}
\eea
where $d_{0}=1$, $d_{1}=2$, $d_{2}=8/5$, 
$$
k_{\gamma} = |\vec{k}| = \frac{M_{H}^{2}-M_{H'}^{2}}{2M_{H}}
\,,
$$
and the anomalous magnetic moment of the heavy quark reads
\be
{\color{black}\kappa_{}=\kappa^{(1)}\alpha_{s}(m)+\kappa^{(2)}\alpha^2_{s}(m)}
\,.
\end{equation}
Note that the use of the effective field theory makes evident that the anomalous magnetic moment 
$\kappa=c^{\rm em}_F -1$ does not have nonperturbative effects. The modification with respect the expressions
deduced in \cite{Brambilla:2005zw} was that the matrix elements were computed using the exact solution of Eq. (\ref{H0}), the use of $\kappa$ to ${\cal O}(\als^2)$, and the explicit implementation of the renormalon cancellation (in particular this is important to get a convergent expansion in $\kappa$).

\begin{table}[hb]
\begin{center}
$$
\begin{array}{|l|c|c|c|c|c|c|}
\hline
   &{\rm LO}&{\cal O}(\als)&{\cal O}(\als^2)&{\cal O}(v^2)&\als \times {\cal O}(\als^2)&v\times{\cal O}(v^2)\\
\hline
\delta \Gamma_{b\bar b} \;({\rm eV})& 14.87 & 1.29 & 0.73 & -1.71& 0.15 &-0.45\\
\hline
\delta \Gamma_{c\bar c} \;({\rm keV}) &2.34 &  0.33 & 0.16 & -0.71 & 0.05 & -0.30\\
\hline
\end{array}
$$ 
\end{center}
\caption{\label{tab:Gammab1S} \it The leading and subleading contributions to $\Gamma_{\Upsilon(1S) \rightarrow \eta_b(1S)\gamma}$ (2nd row) and $\Gamma_{J/\psi (1S) \rightarrow \eta_c(1S)\gamma}$ (third row). 
The last two numbers are error estimates obtained by multiplying the subleading ${\cal O}(\als^2)$ contribution by $\als$ 
and the subleading ${\cal O}(v^2)$ contribution by $v$.}
\end{table}

Eq. (\ref{nS=nS}) was applied to the bottomonium and charmonium ground states. In Table \ref{tab:Gammab1S}
we show the size of the leading and subleading contributions to the decays, as well as some error 
estimates. This gives a flavour of the convergence of the velocity and $\als$ expansion. 
A detailed analysis can be found in Ref. \cite{Pineda:2013lta}. Out of this the following predictions were obtained
for the decays of the bottomonium and charmonium ground states:
\be
\Gamma_{\Upsilon(1S) \rightarrow \eta_b(1S)\gamma}
=15.18 \pm 0.45({\cal O}(v^3)){}^{-0.12}_{-0.05}(N_m){}^{-0.04}_{+0.03}(\als){}^{-0.20}_{+0.20}(m_{\MS})  \;{\rm eV}
\,,
\end{equation}
which after combining the errors in quadrature reads
\be
\label{Gamma1Setab}
\Gamma_{\Upsilon(1S) \rightarrow \eta_b(1S)\gamma}
=15.18(51) \;{\rm eV}
\,;
\end{equation}
\be
\Gamma_{J/\psi(1S) \rightarrow \eta_{c}(1S)\gamma}
=2.12\pm 0.30({\cal O}(v^3)){}^{+0.21}_{-0.23}(N_m){}^{-0.02}_{+0.02}(\als){}^{-0.10}_{+0.11}(m_{\MS}) 
\;{\rm keV}
\,,
\end{equation}
which, after combining the errors in quadrature, reads
 \be
\Gamma_{J/\psi(1S) \rightarrow \eta_{c}(1S)\gamma}
=2.12(40)\;{\rm keV}
\,.
\end{equation}


In Ref. \cite{Pineda:2013lta} the following expressions were used for the hindered transitions
\bea
\label{nVnot=nP}
\Gamma(n^{3}S_{1}\to n'^{1}S_{0}\gamma) &\stackrel{n\neq n'}{=}&
\frac{4}{3}\alpha e_Q^{2} \frac{k_{\gamma}^{3}}{m^{2}} \left[
\frac{k_{\gamma}^{2}}{24}{\color{black}{}_{n'}\langle {r}^{2} \rangle_{n}}+\frac{5}{6}\frac{{\color{black}{}_{n'}\langle {p}^{2} \rangle_{n}}}{m^
{2 } } -
\frac{2}{m^{2}}{\color{black}\frac{{}_{n'}\langle V_{S^2}
(\vec{r}) \rangle_{n}}{E_{n}-E_{n'}}}\right]^{2}, \\
\label{nPnot=nV}
\Gamma(n^{1}S_{0}\to n'^{3}S_{1}\gamma) &\stackrel{n\neq n'}{=}& 4\alpha
e_Q^{2} \frac{k_{\gamma}^{3}}{m^{2}} \left[
\frac{k_{\gamma}^{2}}{24}{\color{black}{}_{n'}\langle {r}^{2} \rangle_{n}}+\frac{5}{6}\frac{{\color{black}{}_{n'}\langle {p}^{2} \rangle_{n}}}{m^
{2 } } 
+
\frac{2}{m^{2}}{\color{black}\frac{{}_{n'}\langle V_{S^2}
(\vec{r}) \rangle_{n}}{E_{n}-E_{n'}}}\right]^{2}, 
\eea
where
\begin{equation}
\nn
V_{S^2}(\vec{r}\,) = \frac{4}{3} \pi C_{f} D_{S^{2},s}^{(2)}(\nu)
\delta^{(3)}(\vec{r}\,)
\,,
\end{equation}
and
$$
D_{S^{2},s}^{(2)}(\nu) = \alpha_{s}(\nu)c_{F}^{2}(\nu)-\frac{3}{2\pi
C_{f}}\left(d_{sv}(\nu)+C_{f}d_{vv}(\nu)\right)
$$
depends on the NRQCD Wilson coefficients. With LL accuracy they read
\bea
\nn
c_{F}(\nu) &= z^{-C_{A}}, \qquad
d_{sv}(\nu) = d_{sv}(m), \\
d_{vv}(\nu) &=
d_{vv}(m)+\frac{C_{A}}{\beta_{0}-2C_{A}}\pi\alpha_{s}(m)(z^{\beta_{0}-2C_{A}}-1)
, 
\eea
\bea
\nn
z &=&
\left[\frac{\alpha_{s}(\nu)}{\alpha_{s}(m)}\right]^{\frac{1}{\beta_{0}}}
\simeq 1-\frac{1}{2\pi}\alpha_{s}(\nu)\ln\left(\frac{\nu}{m}\right), \\
d_{sv}(m) &=& C_{f}\left(C_{f}-\frac{C_{A}}{2}\right)\pi\alpha_{s}(m), 
\qquad d_{vv}(m) = -\left(C_{f}-\frac{C_{A}}{2}\right)\pi\alpha_{s}(m).
\eea

\begin{figure}[ht]
\begin{center}
\epsfxsize=4.65in
\epsffile{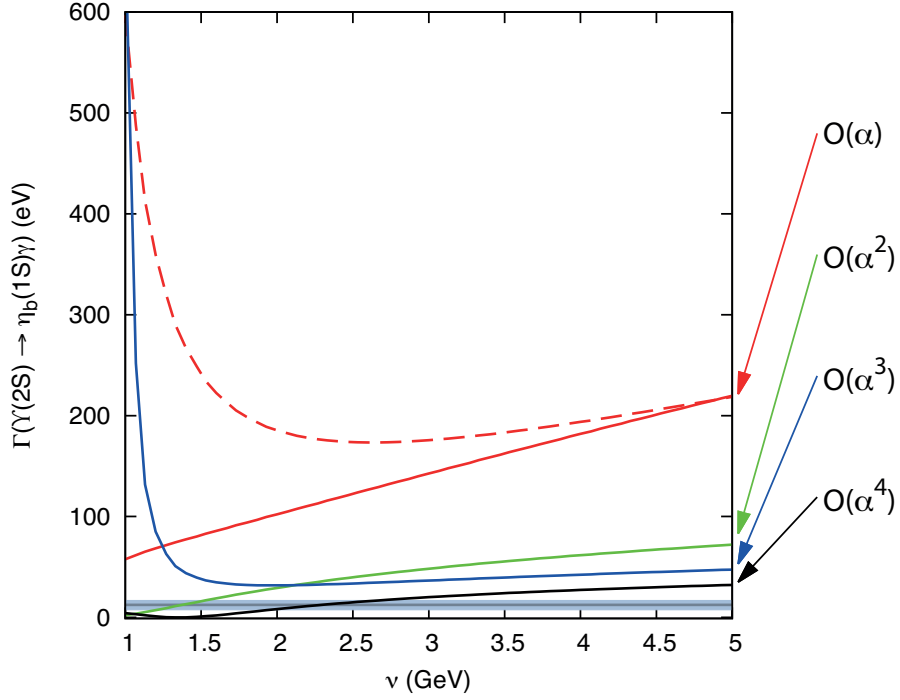}
\end{center}
\caption{\label{fig:botn2decay} \it Plot of $\Gamma_{\Upsilon(2S) \rightarrow \eta_b(1S)\gamma}$ using the static potential $V_{\RS'}^{(N)}$ at different orders in perturbation theory: $N=0,1,2,3$. The dashed line corresponds to no resumming the hard logarithms: $D_{S^2,s}=\als(\nu)$. The blue band corresponds to the experimental value \cite{Beringer:1900zz}. From \cite{Pineda:2013lta}.
}
\end{figure}

The modification with respect the expressions
deduced in \cite{Brambilla:2005zw} was that the matrix elements were computed using the exact solution of Eq. (\ref{H0}), the resummation of the hard logarithms in $V_{S^2}$, and the  implementation of the renormalon cancellation when necessary.

\section{E1 transitions}

In Ref. \cite{Brambilla:2012be} theoretical expressions for the E1 transitions were obtained both in the weak and strong coupling version of pNRQCD, which can be summarized in the following expression
\bea
 \label{eq:E1}
 \Gamma _{n^3 P_J \rightarrow {n'}^3 S_1 \gamma} &=& \frac{4}{9} \, \alpha _{em} e_Q ^2 k_{\gamma}^3 I_3 ^2(n1 \rightarrow n'0) 
 \\
 &&
 \times\left( 1 + R - \frac{k_{\gamma}^2}{60} \frac {I_5}{I_3} - \frac{k_{\gamma}}{6m} + \frac{\kappa 
 k_{\gamma} }{2m} \left[\frac{J(J+1)}{2} -2 \right] \right) \, ,
\nn
\eea
where
\begin{equation}
\nn
  I_s \equiv \int _0 ^{\infty} dr  \, r^s R_{n'0} (r) R _{n1} (r) \, .
\end{equation}
$R$ contains all the wave-function corrections due to higher-order potentials, the relativistic correction of the kinetic energy, $-p^4/4m^3$, and higher-order Fock state contributions due to intermediate color-octet states (we refer to \cite{Brambilla:2012be} for detailed expressions). In contrast to M1 transitions the latter ones do
not vanish for E1 decays.
Eq.~(\ref{eq:E1}) (without color-octet contributions
in R) is also valid in the strongly coupled regime.
The expressions obtained with potential models in \cite{Grotch:1984gf} miss the color-octet
contributions in the weak-coupling regime and the contributions coming from the $1/m$ potential at strong coupling. 
Without much effort one can extend the discussion to other processes like $n{}^1P_1 \rightarrow n'{}^1S_0\gamma$
and $n{}^3S_1 \rightarrow n^{\prime}{}^3P_J\gamma$, also for transitions between states with the same principal quantum
number, where corrections $ \sim k_\gamma$ are suppressed.

\section{Conclusions}

We have reviewed recent model independent determinations of heavy quarkonium 
radiative transitions in pNRQCD. For the magnetic dipole transitions the precision reached was
$k_{\gamma}^3/m^2\times{\cal O}(\als^2,v^2)$ and $k_{\gamma}^3/m^2\times{\cal O}(v^4)$ for the allowed and forbidden transitions, respectively. Large logarithms associated with the heavy quark mass scale were also resummed. The effect of the improved power counting was found to be large, and the exact treatment of the soft logarithms of the static potential made the factorization scale dependence much smaller. The convergence for the $b\bar b$ ground state was quite good, and also quite reasonable for the $c\bar c$ ground state and the $b\bar b$ $1P$ state. For all of them solid predictions were given, which we summarize here \cite{Pineda:2013lta}:
\bea
\Gamma_{\Upsilon(1S) \rightarrow \eta_b(1S)\gamma}
&=&15.18(51) \;{\rm eV}
\,,
\\
\Gamma_{J/\psi(1S) \rightarrow \eta_{c}(1S)\gamma}
&=&2.12(40)\;{\rm keV}
\,,
\\
\Gamma_{h_b(1P) \rightarrow \chi_{b0}(1P)\gamma}
&=&0.962(35)\;{\rm eV}
\,,
\\
\Gamma_{h_b(1P) \rightarrow \chi_{b1}(1P)\gamma}
&=&8.99(55)\times 10^{-3}\;{\rm eV}
\,,
\\
\Gamma_{\chi_{b2}(1P) \rightarrow h_b(1P)\gamma}
&=&0.118(6)\;{\rm eV}
\,.
\eea

For the $2S$ decays the situation is less conclusive. The ${\cal O}(v^2)$ correction of the $\Upsilon(2S) \to \eta_b(2S)\,\gamma$ decay suffered from a bad convergence in $N$, producing relatively large errors for the prediction. Some of the ${\cal O}(v^2)$ matrix elements of the $\eta_b(2S) \to \Upsilon(1S)\,\gamma$ decay also suffered from this bad convergence. This impedes giving a reliable error estimate for this transition, as such terms correspond to the leading (and only known so far) order expression (moreover, they should be squared in the decay). The situation is completely different for the $\Upsilon(2S) \to \eta_b(1S)\,\gamma$ transition. The reason is that the problematic 
${\cal O}(v^2)$ matrix elements appear in a different combination for this decay, so that they cancel to a large extent. This led to a nicely convergent sequence in $N$ (as we illustrate in Fig. \ref{fig:botn2decay}), where the resummation of the hard logarithms played an important role. The final figure was \cite{Pineda:2013lta}
\be
 \Gamma^{(\rm th)}_{\Upsilon(2S) \rightarrow \eta_b(1S)\gamma}
=6^{+26}_{-06} \, {\rm eV}.
\end{equation}
This number is perfectly consistent with existing data, so that the previous disagreement with experiment for the $\Upsilon(2S) \to \eta_b(1S)\,\gamma$ decay fades away. 

For the M1 transitions of the low lying heavy quarkonium states discussed above a pure weak coupling analysis was suitable and nonperturbative effects subleading. This is not so for the E1 transitions. Non-perturbative effects start to appear at ${\cal O}(v^2)$ in the weak coupling version of pNRQCD. 
For the strong coupling limit the result depends on wave functions and expectation values of operators obtained with non-perturbative potentials. 
These non-perturbative effects may introduce large uncertanties to phenomenological applications of 
Eq. (\ref{eq:E1}) and calls for a dedicate study of them. Nevertheless, 
in the mean time, there have been some preliminary phenomenological analysis \cite{Pietrulewicz:2013ct} neglecting octet effects, and using some parameterizations of the potentials aiming to merge perturbation theory 
at short distances with (when possible) string models at long distances.
The results are very encouraging getting good agreement with experiment when applied to a large variety of decays (see Tables \ref{table_bottomonium} and \ref{table_charmonium}). 
   
\medskip

{\bf Acknowledgements}.
I am grateful to Jorge Segovia for collaboration on part of the work reviewed here. 

\begin{table}
\centering
   \begin{tabular}{c|c|c|c|c}
	process & $\Gamma_{\rm{pNRQCD}}^{\rm{LO}}$/keV & $\Gamma_{\rm{pNRQCD}}^{\rm{NLO}}$/keV & $\Gamma_{\rm{mod}}$/keV & $\Gamma_{\rm{exp}}^{\textrm{PDG}}$/keV \\
	\hline
	$\chi_{b0}(1P) \rightarrow \Upsilon(1S) \gamma$ & 31.8 & 29.7 $\pm$ 3.1 & 25.7-27.0 & - \\ 
	$\chi_{b1}(1P) \rightarrow \Upsilon(1S) \gamma$ & 40.3 & 35.8 $\pm$ 4.0 & 29.8-31.2 & - \\ 
	$\chi_{b2}(1P) \rightarrow \Upsilon(1S) \gamma$ & 45.9 & 40.6 $\pm$ 4.6 & 33.0-34.2 & - \\ 
	$h_b(1P) \rightarrow \eta_b(1S) \gamma$ & 60.8 & 44.3 $\pm$ 6.1 & - & - \\
	$\Upsilon(2S) \rightarrow \chi_{b0}(1P) \gamma$ & 1.52 & 1.13 $\pm$ 0.15 & 0.72-0.73 & 1.22 $\pm$ 0.16 \\
	$\Upsilon(2S) \rightarrow \chi_{b1}(1P) \gamma$ & 2.26 & 1.94 $\pm$ 0.23 & 1.62-1.65 & 2.21 $\pm$ 0.22 \\
	$\Upsilon(2S) \rightarrow \chi_{b2}(1P) \gamma$ & 2.34 & 2.19 $\pm$ 0.23 & 1.84-1.93 & 2.29 $\pm$ 0.22 \\
	$\chi_{b0}(2P) \rightarrow \Upsilon(2S) \gamma$ & 12.6 & 13.0 $\pm$ 1.3 & 10.6-11.4 & - \\
	$\chi_{b1}(2P) \rightarrow \Upsilon(2S) \gamma$ & 17.1 & 16.3 $\pm$ 1.7 & 11.9-12.5 & - \\
	$\chi_{b2}(2P) \rightarrow \Upsilon(2S) \gamma$ & 20.4 & 18.1 $\pm$ 2.0 & 12.9-13.1 & - \\
	$\Upsilon(3S) \rightarrow \chi_{b0}(2P) \gamma$ & 1.44 & 1.05 $\pm$ 0.14 & 1.07-1.09 & 1.20 $\pm$ 0.16 \\
	$\Upsilon(3S) \rightarrow \chi_{b1}(2P) \gamma$ & 2.38 & 2.05 $\pm$ 0.24 & 2.15-2.24 & 2.56 $\pm$ 0.34 \\
	$\Upsilon(3S) \rightarrow \chi_{b2}(2P) \gamma$ & 2.53 & 2.35 $\pm$ 0.25 & 2.29-2.44 & 2.66 $\pm$ 0.41 
  \end{tabular} 
  \caption{\it E1 decay rates for bottomonium. pNRQCD results compared to a potential model calculation \cite{Grotch:1984gf}  and and the current PDG values \cite{Beringer:1900zz}. LO denotes the result obtained without relativistic corrections, NLO indicates the result up to $\mathcal{O}(v^2)$ neglecting color-octet effects in the weak-coupling regime and non-perturbative contributions to $V_r^{(2)}$. The error estimates give the generic size of one $\mathcal{O}(v^2)$ correction as well as an estimate for the sum of all corrections at $\mathcal{O}(v^3)$. From \cite{Pietrulewicz:2013ct}.}
  \label{table_bottomonium}
\end{table}
\begin{table}
\centering
 \begin{tabular}{c|c|c|c|c}
	process & $\Gamma_{\rm{pNRQCD}}^{\textrm{LO}}$/keV & $\Gamma_{\rm{pNRQCD}}^{\textrm{NLO}}$/keV & $\Gamma_{\rm{mod}}$/keV & $\Gamma_{\rm{exp}}^{\textrm{PDG}}$/keV \\
	\hline
	$\chi_{c0}(1P) \rightarrow J/\psi \gamma$ & 199 & 158 $\pm$ 60 & 162-183  & 122 $\pm$ 11 \\
	$\chi_{c1}(1P) \rightarrow J/\psi \gamma$ & 421 & 302 $\pm$ 126 & 340-363  & 296 $\pm$ 22  \\
	$\chi_{c2}(1P) \rightarrow J/\psi \gamma$ & 568 & 415 $\pm$ 170 & 413-464 & 386 $\pm$ 27 \\
        $h_c (1P) \rightarrow \eta_c(1S) \gamma$ & 909 & 447 $\pm$ 272 & - & $<$600  \\
        $\psi(2S) \rightarrow \chi_{c0}(1P) \gamma$ & 53.6 & 21.4 $\pm$ 16.1 & 26.0-40.3  & 29.4 $\pm$ 1.3 \\
	$\psi(2S) \rightarrow \chi_{c1}(1P) \gamma$ & 45.2 & 30.7 $\pm$ 13.6 & 28.3-37.3  & 28.0 $\pm$ 1.5  \\
	$\psi(2S) \rightarrow \chi_{c2}(1P) \gamma$ & 31.6 & 25.6 $\pm$ 9.5 & 17.5-22.7 & 26.5 $\pm$ 1.3 \\
        $\eta_c(2S) \rightarrow h_c(1P) \gamma$ & 38.1 & 31.0 $\pm$ 11.4 & - & -
  \end{tabular} 
  \caption{\it E1 decay rates for charmonium.  pNRQCD results at LO, NLO (including error estimate) compared to a potential model calculation \cite{Grotch:1984gf} and the current PDG values \cite{Beringer:1900zz}. From \cite{Pietrulewicz:2013ct}.}
  \label{table_charmonium}
\end{table}

\end{document}